\documentclass{aastex}

\shorttitle{Structure of the SMC}
\shortauthors{Crowl, et al.}

\begin{document}

\title{The Line-of-Sight Depth of Populous Clusters in the Small
  Magellanic Cloud}
\author{Hugh H. Crowl\altaffilmark{1} and Ata
  Sarajedini\altaffilmark{2,3}}
\affil{Astronomy Department, Wesleyan University, Middletown, CT 06459}
\email{hugh@astro.yale.edu; ata@astro.ufl.edu}

\altaffiltext{1}{Current Address: Astronomy Department, Yale University,
P. O. Box 208101, New Haven, CT 06520}

\altaffiltext{2}{Guest User, Canadian 
    Astronomy Data Centre, which is operated by the Dominion
    Astrophysical Observatory for the National Research Council of
    Canada's Herzberg Institute of Astrophysics.}

\altaffiltext{3}{Current address: University of Florida, Department of
  Astronomy, Gainesville, FL 32611}

\author{Andr\'{e}s E. Piatti}
\affil{Observatorio Astro\'omico, Laprida 854, 5000, C\'ordoba, 
Argentina}
\email{andres@mail.oac.uncor.edu}

\author{Doug Geisler}
\affil{Grupo de Astronomia, Departmento de Fisica, Universidad de
Concepci\'{o}n, Casilla
  160-C, Concepci\'{o}n, Chile}
\email{doug@kukita.cfm.udec.cl}

\author{Eduardo Bica}
\affil{Departamento de Astronomia, Instituto de Fisica, UFRGS, CP
  15051, 91501-970, Porto Alegre, RS, Brazil}
\email{bica@if.ufrgs.br}

\author{Juan J. Clari\'{a}}
\affil{Observatorio Astro\'omico, Laprida 854, 5000, C\'ordoba, 
Argentina}
\email{claria@mail.oac.uncor.edu}

\and

\author{Jo\~{a}o F. C. Santos, Jr.}
\affil{Departamento de Fisica, ICEx, Universidade Federal de Minas 
  Gerais, CP 702, 30123-970 Belo Horizonte, MG, Brazil}
\email{jsantos@fisica.ufmg.br}

\begin{abstract}
We present an analysis of age, metal abundance, and positional
data on populous clusters in the Small Magellanic Cloud (SMC) with
the ultimate aim of determining the line-of-sight (LOS) depth of the SMC
using these clusters as proxies. Our dataset contains 12 objects and
is limited to clusters with the highest quality data for which the ages 
and abundances are best known and can be placed on an internally 
consistent scale. We have analyzed the variation of the clusters' 
properties with position on the sky and with line-of-sight depth. 
Based on this analysis, we draw the following conclusions.
1) The observational data indicates that the eastern side
of the SMC (facing the LMC) contains younger and more metal-rich
clusters as compared with the western side. This is not a strong
correlation because our dataset of clusters is necessarily limited,
but it is suggestive and warrants further study.
2) Depending on how the reddening is computed to our clusters, we find
a mean distance modulus that ranges from  $(m-M)_0 = 18.71 \pm 0.06$
to $(m-M)_0 = 18.82 \pm 0.05$.
3) The intrinsic $\pm$1-$\sigma$ line-of-sight depth of the SMC
populous clusters in our study is between \(\sim 6\) kpc and
\(\sim 12\) kpc depending primarily on whether we adopt the Burstein \&
Heiles
reddenings or those from Schlegel et al. 
4) Viewing the SMC as a triaxial galaxy with the Declination, 
Right Ascension, and LOS depth as the three axes, we find axial ratios 
of approximately 1:2:4.
Taken together, these conclusions largely agree with those of
previous investigators and serve to underscore the utility of populous
star clusters as probes of the structure of the Small Magellanic Cloud.
\end{abstract}


\section{Introduction}
\label{intro}
The effect of interactions/mergers on the star formation histories
of galaxies has become a field of intense activity 
(e.g. Carlson et al. 1998; Whitmore et al. 1999; Elmegreen et al. 2000). 
While many investigators
have studied distant galaxy systems as a means of addressing this
question, the Milky Way / Large Magellanic / Small Magellanic Cloud system
provides a nearby example
that is especially profitable scientifically (e.g. Gardiner 1999). This is
because the relative proximity of these galaxies allows us to perform
detailed age and abundance studies of individual field and cluster
stars. Given the possibility that the Milky Way and other spiral
galaxies formed via the accretion/merger of dwarf satellite galaxies
\citep{searle78, cote00},
understanding the interplay between dynamical interactions and star
formation is important in unlocking the secrets of galaxy formation. 

One of the first pieces of evidence for the 
occurrence of dynamical interactions in the Milky Way/LMC/SMC
system was the discovery of the Magellanic Stream by \citet{wannier72}. 
\citet{welch87}, \citet{westerlund90}, \citet{bica95} and \cite{bica99} 
discuss evidence for interactions between the Clouds based on the 
spatial distributions of
stellar population tracers, such as the star clusters, associations, 
and emission nebulae. The recent work of
\citet{kunkel00} reviews the formation of the Magellanic Stream and the
interactions between the Large and Small Magellanic Clouds in
detail. There is a fair amount of evidence, both from simulations
\citep{murai80} and observations 
\citep{mathewson88,hatz89a, gardiner91} that, as
a result of these interactions, the
SMC may extend beyond its tidal radius. Some authors \citep{mathewson84,
mathewson86} have even suggested that the SMC has split into two
components: the Mini Magellanic Cloud (MMC) and the SMC Remnant. In a
contrasting view, \citet{welch87} utilize infrared observations
of cepheids to assert that the spatial extent of the SMC is well
within its tidal radius. One
of the most important pieces of evidence in determining the validity
of these claims is the overall line of sight (LOS) depth of the Small
Magellanic Cloud (SMC).

The most extensive series of papers dealing with the LOS depth of the
SMC are those by \citet{hatz89a}, \citet{hatz89b}, \citet{gardiner91},
and \citet{gardiner92}. These authors studied the depth of the SMC 
using the magnitude spread of the horizontal branch/red clump 
(HB/RC) stars among the
SMC {\it field} population. Their photographic survey, augmented by CCD
observations of select fields, was quite extensive, covering 48.5 
square degrees centered on the SMC. The basic premise of their
investigation was that the magnitude of the HB/RC stars is mainly
affected by distance and photometric errors but minimally influenced
by the physical properties of the stars themselves, such as metallicity
and age. Based on this assumption, they conclude that, not only is
the SMC significantly dispersed in the LOS direction, but that the
amount of dispersion varies with position on the sky.
For example, the northeastern region located more than
2 kpc from the optical center suffers the greatest LOS dispersion with 
an average LOS depth of 17 kpc and a maximum of 23 kpc. 
In contrast, the analogous region located
in the southwest displays a depth that is $\sim$10 kpc shallower on
average. In addition, most of the areas to the north and northwest of 
the optical center display a depth of between 4 and 9 kpc.

In the years since these landmark studies, there has been some
controversy about the age and metallicity sensitivity of the
red clump absolute magnitude. The reader is referred to the papers 
by \citet{paczynski98},
\citet{udalski98}, \citet{cole98}, \citet{sarajedini99}, 
and \cite{salaris00} for some perspective on this question.
The degree of sensitivity could have a significant effect on the derived
LOS depth of the SMC based on the magnitude of the red clump.
One way to explore the extent of this effect is to use observations
of red clumps in SMC star clusters where we can define the age
and metallicity far more precisely than in field star populations.

In order to explore the LOS depth of the SMC and the effect of 
assumptions concerning the variation of the red clump absolute 
magnitude, we have studied the distances to twelve SMC
populous clusters that possess 
red clumps. These clusters are
located all across the face of the SMC and span over three
degrees in declination and almost one hour in right ascension. There
are also several clusters near the right ascension of the optical center
of
the SMC (\(\alpha_{2000}= 0^h 52^m 45^s, \delta_{2000}=- 72\degr 49'
43''\)). We will determine distances to these clusters assuming first that
the sensitivity of the RC to age and metallicity is negligible,
and then that the sensitivity is significant and is well-described 
by theoretical models. In Section 2, we will outline the sources from
which
the data were drawn and discuss the observational quantities obtained
from the data. Section 3 is concerned with the theoretical isochrones and
their calibration. Section 4 presents a discussion of our findings with
regard to the depth of the SMC, as well as correlations between various
cluster parameters. Finally, Section~5 presents the results and
conclusions.

\section{Observational Data}
\label{obs}

The observational data for this study, twelve SMC populous
clusters spanning a range of ages and metallicities but all possessing
predominantly
 red horizontal branches, are of three different pedigrees. 
Seven of the clusters (Lindsay 113, Kron 3, NGC 339, NGC 416, NGC 361,
Lindsay 1, NGC 121) are taken from \citet{mighell98}. In that work, 
these seven intermediate metallicity 
clusters (\(-1.7 \lesssim [Fe/H] \lesssim -1.2\)) were
studied in order to better understand the star formation and
chemical enrichment history of the Small Magellanic Cloud. 
Three of the clusters (Kron 28, Lindsay 38, Kron 44)
are taken from the paper by \citet{piatti01}. These data were obtained in
the $C$ and $T_1$ filters of the
Washington photometric system using the 0.9m telescope at Cerro
Tololo Inter-American Observatory. The observational data for the
remaining two clusters (NGC 411 and
NGC 152) come from archival Hubble Space Telescope /
Wide Field Planetary Camera 2 images. These observations,
taken as part of a snapshot survey of LMC/SMC clusters \citep{rich00},
consist of one image in the F450W ($\sim$B) filter  and another in the
F555W ($\sim$V) filter for each cluster.\footnote{We had hoped to add NGC
419
to our dataset of SMC clusters. However, the CMD of `NGC 419' presented
by \citet{rich00} is actually that of NGC 411 observed at a different
roll angle than the ostensive NGC 411 CMD. It is therefore not 
surprising that \citet{rich00} find NGC 411 and `NGC 419' to have
identical ages.}  The observational data for these clusters are listed in
Table~\ref{table:HSTobs}. 
In order to visualize the positions of the clusters
within the field population of the 
SMC, Figure 1 shows a mosaic of Digitized Sky Survey images downloaded 
from Skyview\footnote{SkyView
  was developed and is maintained under NASA ADP Grant NAS5-32068 with
  P.I. Thomas A. McGlynn under the auspices of the High Energy
Astrophysics
  Science Archives Research Center (HEASARC) at the GSFC Laboratory
  for High Energy Astrophysics.}. This figure illustrates that, even
though our
dataset of clusters is relatively small, they cover a substantial
portion of the SMC's spatial extent. We note that our cluster
database is limited by the available resources. In particular,
there is a serious dearth of published CCD 
observations for SMC star clusters. Nonetheless, as the database of SMC 
cluster observations increases, the techniques we 
have utilized in the present work will 
continue to be important as a probe of the SMC's three-dimensional
structure.

\subsection{Reduction of HST Observations}

The standard image processing steps (de-biasing, 
flat-fielding, creating a bad pixel mask, etc) were
performed by the Canadian Astronomy Data Centre\footnote{Canadian 
Astronomy Data Centre is operated by the Dominion Astrophysical
Observatory for the National Research Council of Canada's Herzberg
Institute of Astrophysics.}. The data reduction procedure was similar to
that
employed by \citet{sarajedini98} and the reader is referred to that paper
for more details.\footnote{While these observations
  were reduced in the work of \citet{rich00}, one of the most
  important concerns in the present study is internal consistancy, so we
  have re-reduced these images in the same way as the other HST clusters.
Note
  that our results do not vary significantly from those of
  \citet{rich00}.}  Using DAOPHOT II \citep{stetson94}, we measured the
magnitude of all detected profiles on each CCD (PC1, WF2, WF3, and WF4)
within a radius of 2 pixels.
Aperture corrections to a 0.5'' radius (10.87 PC1 pixels, 5 WF pixels)
were then determined using 30 to 50 of the brightest uncrowded stars 
on each frame. With the aperture corrections applied, the F450W and F555W
photometry was
matched to form colors and transformed to the standard system of
\citet{holtzman95}. A 12\% charge transfer efficiency correction in the
y-axis direction was applied following the recommendations of
\citet{holtzman95}
for observations taken at an operating temperature of $-76$\degr C. Using
this reduction procedure, \citet{sarajedini98} found good agreement
between his HST/WFPC2 photometry for the LMC cluster NGC 2193 and
ground-based photometry for the same cluster from \citet{dkm}.

\subsection{Red Clump Magnitude}

In the case of the Mighell et al. (1998) clusters, the peak V magnitude 
of the red clump has already been computed and appears in their Table 9.
\footnote{It should be noted that the value of $I_0(RC)$ for Lindsay 
113 is incorrect in Table 5 of \citet{udalski98}. Instead of 18.33,
the value of $I_0(RC)$ should be 18.13.}
For the remaining two sets of data (Piatti et al. 2001 and the HST
photometry), color magnitude diagrams were first constructed as
shown in Figures 2 through 5. 
On each of these diagrams, a box was drawn encompassing the location
of the red clump in each cluster. The median $V$ magnitude of the stars in 
that box was taken to be representative of the peak
$V$ magnitude of the red clump
($V(RC)$). The median was used to minimize the effect of outliers
on $V(RC)$. Errors in the red clump magnitude were
calculated from the standard error about the median. For NGC 411
and NGC 152, an additional error of 0.05 mag has been added 
due to uncertainties in the absolute photometric scale (see 
Mighell et al. 1998). These apparent
magnitudes are provided in Table 2.

In the case of NGC 411 and NGC 152, there may be a concern 
that the measured red clump magnitude of the cluster
has been unduly influenced by the red clump of the SMC field surrounding
each cluster. Keeping in mind that
the cluster was centered in the planetary camera (PC1), it is reasonable
to assume that the cluster stars, and thus the cluster red clump,
will be most prominent in the PC1 image. Indeed, we find this to be the
case both for NGC 152 
(Figure 3) and NGC 411 (Figure 4), where more red clump stars per unit
area\footnote{Since the planetary camera has $\sim$~25\% the field of
view of the other cameras, with all other things being
equal, one would expect $\sim$~25\% as many stars.} are present in the 
planetary camera as compared with the wide field chips. 
We find that the value of $V(RC)$ for the 
wide field cameras is identical to that of PC1. Therefore, we can safely 
combine the CMDs of all of the frames (Figure 5) and determine $V(RC)$
from the composite CMD.  

As mentioned above, the observations of \citet{piatti01} were obtained 
using the Washington \citep{canterna76} photometric system. The
Washington system consists of four different filters, some of which
roughly approximate those in the Johnson-Kron-Cousins UBVRI
system. \citet{geisler96} has provided robust relations that
can be used to convert between these two systems.
We begin by transforming the $T_1$ magnitude of the red clump
to Johnson $V$
using the coefficients in Table 4 of \citet{geisler96}. 
In order to do this, it was
first necessary to transform the RC $T_1$ magnitude into an $R$
magnitude. From there, we converted the $C-T_1$ color into a $V-R$
color. Having $R$ and $V-R$, we then simply determined the 
$V$ magnitude of the red clump. The errors in $V(RC)$ were calculated by
transforming \(T_1+1\)\(\sigma_{T_1}\) and 
\(T_1-1\)\(\sigma_{T_1}\) into \(V\) magnitudes and using
those values to deduce the standard error about the median in
\(V\). In addition to these standard errors, there
was an uncertainty introduced when the transformation was
applied. These errors, which are presented in Table 4 of
\citet{geisler96}, were added in quadrature to the standard errors
about the median to obtain the total uncertainty in $V(RC)$. 

We note that the clusters K28, L38, and K44 are located near the optical
center of the SMC where the field star contamination is high. As a result, 
we have used the field subtracted photometry for these clusters derived by
Piatti et al. (2001) and plotted in Figure 2. We have used the median to 
compute the red clump magnitude because it is relatively insensitive to
which subset of cluster stars are used to locate the red clump.

\subsection{Interstellar Extinction}

We seek an approach that yields internally consistent reddenings for
each of our clusters. As such, we have avoided using individual literature 
values for each cluster opting instead to utilize the reddening maps 
of \citet{bh} and \citet{schlegel98}. First, the traditional \citet{bh} 
(hereafter BH) reddening maps 
are based primarily on HI column densities at declinations appropriate
for the SMC. Second, the more modern COBE/DIRBE reddening maps of 
\citet{schlegel98} (hereafter SFD) provide all-sky extinction values
determined from the infrared emission of interstellar dust. 
In this study, we will make use of both of these options making sure to
point out significant differences in the analysis as they occur.
To determine $A_V$, we
have adopted \(A_V=3.1 \cdot E(B-V)\) \citep{cardelli89,schlegel98}. 
Since no formal errors are given 
by the maps, the errors in the reddening are taken to be 10\%
of the reddening value itself.

\subsection{Ages and Metallicities}

Because the cluster age and metallicity could influence the luminosity of
the red clump \citep{sarajedini99,girardi00}, we also require 
knowledge of 
these quantities in order to determine the absolute red clump magnitude 
($M_V(RC)$) and thus the distance to each cluster.
Published values were adopted for these quantities as given in 
Table 2. The majority are from the
previously mentioned \citet{mighell98} work, where a large amount of 
our photometric data originated. The metallicities for these clusters
were determined using the simultaneous reddening and metallicity (SRM)
method of \citet{sarajedini94} as well as a new calibration of the
RGB slope as a function of metallicity, both of which are on the
\citet{zw84} abundance scale. 
The ages of the clusters in \citet{mighell98} were determined via 
the technique described by \citet{sll95}, which uses the difference 
between the RGB color and the age-sensitive HB color, corroborated by
isochrone comparisons to the main sequence turnoff 
regions.\footnote{Mighell et al. (1998) correctly point out that age
estimates for their clusters based {\it only} on the main sequence turnoff
would be subject to significant uncertainty. This is because the
short exposure archival HST observations on which their 
study is based result in unacceptably large photometric
scatter in the region of the turnoff. If Mighell et al. had
relied solely on the magnitude and color of the turnoff to
determine the cluster ages, they would not have attained the
high level of precision that ultimately resulted from their analysis.} 
Their age scale assumes that Lindsay 1 is 9 Gyr old 
\citep{osm96}. 

In the case of the three clusters from the work of \citet{piatti01},
the metal abundances are a result of comparing
the observed RGBs of the clusters with the standard RGBs in the
Washington photometric system set up by \citet{geisler99}.
Note that Kron 28  has a different metal abundance depending
on whether \citet{piatti01} adopt the \citet{bh} reddening maps
or those of \cite{schlegel98} (see Section 4).
This is indicated in Table 2 by
listing two numbers with the latter in parenthesis.
The ages of these clusters rely upon the magnitude difference 
between the red clump and
the main sequence turnoff along with a calibration provided by
\citet{geisler97} that relates this magnitude
difference to age. Again, the metal abundances are on the scale
of \citet{zw84}, and the age scale assumes 9 Gyr for Lindsay 1. 

Lastly, we consider the clusters whose photometry was measured 
specifically for the present study. The metallicity
of NGC 411 has been determined in \citet{alves99} using the same SRM
method \citep{sarajedini94} as \citet{mighell98}. Because NGC 411 is
significantly younger than the other clusters, however,
\citet{alves99} use a formulation for Galactic open clusters from
\citet{noriega97}, which yields
an abundance that is in good agreement with other determinations.
To estimate the metal abundance of NGC 152, we utilize the same
technique employed by \citet{alves99} in the case of NGC 411.
We begin by fitting a polynomial to the RGB. This is then input
into the SRM method as modified by \citet{noriega97}, 
which yields a metallicity of 
$[Fe/H] = -0.94 \pm 0.15$ for NGC 152.
The age of NGC 411 was determined by \citet{alves99} using
the \citet{bertelli94} isochrones fitted to the main sequence 
turnoff. They find that NGC 411 is $1.4\pm0.2$ Gyr old.
Using these same isochrones, \citet{mighell98} find an
age of $9\pm1$ Gyr for Lindsay 1. Because our age scale is based
on Lindsay 1 being 9 Gyr old, no correction has been applied to
the age of NGC 411. 
For the age of NGC 152, we again employ the technique of
\citet{alves99} and fit the \citet{bertelli94}
isochrones to our CMD which, like the case of NGC 411, yields an 
age of $1.4\pm0.2$ Gyr for NGC 152.

To close out this section, we seek to compare our cluster ages and
metallicities with recent compilations in the literature. In
particular, for the 7 clusters in common with the study of
\citet{udalski98}, we find $\Delta$$[Fe/H]$$=-0.05 \pm 0.05$ dex
and $\Delta$Age $=+0.1 \pm 0.3$ Gyr, where the differences have been
computed in the sense (Us--Udalski). Similarly, for the 7 clusters
in common with \citet{dacosta98}, we find 
$\Delta$$[Fe/H]$$=-0.04 \pm 0.05$ dex and $\Delta$Age $=-0.5 \pm 0.5$ Gyr. 
These differences are not statistically significant and thus provide 
support for the validity of our cluster ages and abundances.

\section{Theoretical Isochrones}

In order to determine the variation of the red clump
absolute magnitude [$M_V(RC)$] with age and metal abundance, we
utilize the theoretical horizontal branch models of
\citet{girardi00} (see also \citet{salaris00}), which span the 
metallicity range from Z=0.0004
to Z=0.030; additional models with Z=0.0001 were kindly provided
by Leo Girardi. There is now abundant evidence supporting 
the validity of these theoretical models for the purpose of the present
analysis. Not only is there the work 
of \citet{sarajedini99} and \citet{gmc00} which compare the models to open
cluster red clumps, 
but there is also the work of \citet{girardi99} and \citet{salaris00}
which compare them to the red clumps of composite 
populations such as the Galactic bulge and the LMC. In addition,
\citet{castellani00} present a comparison of the 
Girardi et al. models with those of other authors.

Figure 6 shows a spline fit to the median absolute magnitude of the
red clump as a function of metallicity and age. Isochrones for ages of 
(top to bottom) 2.0, 2.5, 5.0, 7.0, 7.9, 10.0, and 12.6 Gyr are shown
by the solid lines. An isochrone for an age of 1 Gyr is shown by the
dashed line.
For ages younger than $\sim$2 Gyr, the isochrones become fainter again
before reversing course and brightening for ages younger than $\sim$1.2 
Gyr. We utilize isochrones as young as 1 Gyr in our calculations. In order
to calibrate the distance scale of the theoretical
models to observational data, we rely upon 
$M_V(RC)$, age, and metallicity values for eight open clusters 
from \citet{sarajedini99}. In that work, the distance modulus to each 
open cluster was determined using main sequence fitting to M67,
which was in turn fitted to the solar abundance \citet{bertelli94}
isochrones. 
Utilizing the distance determined via this method, along with
$V(RC)$, \citet{sarajedini99} was able to compute $M_V(RC)$
for all of the clusters in his study. In the present analysis, we use 
the mean values of the
eight open clusters from \citet{sarajedini99}
to calibrate the theoretical models. The average
red clump absolute magnitude of the open clusters is $M_V(RC)=0.950 \pm
0.035$ at an average age of \(2.68 \pm 0.75\) Gyr and an average
metallicity of \(-0.04 \pm 0.06\) dex as shown in Fig. 6. The theoretical
values of $M_V(RC)$ from Girardi et al. were increased by +0.093 mag to 
bring them onto the distance scale of \citet{sarajedini99}. These 
calibrated models were
then used to determine the absolute magnitude of the red clump for the
twelve clusters in this work as described in the next section.

\section{Results and Discussion}

Using the theoretical isochrones illustrated in Fig. 6, 
we are able to determine the absolute magnitude of the red clump 
($M_V(RC)$) for the twelve clusters in our sample. These absolute 
magnitudes, when coupled with the measured
apparent magnitude of the red clump ($V(RC)$) and the interstellar
extinction ($E(B-V)$), yield the distance to each cluster.
The total error in the distance includes errors in the values
of $M_V(RC)$, $V(RC)$ and \(A_V\). 

The resultant distances are listed in Table 3 under different
assumptions for the variation of $M_V(RC)$ (line 1 and 2 for
each cluster) and the adopted reddenings (columns 3 and 4 
for each cluster).
We consider four distinct scenarios. 1) The absolute magnitude
of the red clump is influenced by age and metallicity, as given in
Fig. 6, coupled with the BH reddening values. 2) This is the same
as (1) except that we adopt the SFD reddenings.
3) The absolute magnitude of the red clump is not influenced by 
age and metallicity, in which case we adopt the mean $M_V(RC)$ of
our 12 clusters ($\langle$$M_V(RC)$$\rangle$$ = 0.56 \pm 0.04$), 
along with the BH reddening values. 4) This is the
same as (3) except that the SFD reddenings are used.

Given these cluster distances computed under various assumptions,
we can calculate the mean distance modulus of our 12 clusters. 
Under assumptions (1) and (3) above, both of which use the BH reddenings, 
we find $(m-M)_0 = 18.82 \pm 0.05$, while assumption (2) and (4)
yield $(m-M)_0 = 18.71 \pm 0.06$ based on the SFD reddenings. The
quoted errors represent standard errors of the mean for the 12 clusters.
It is important to reiterate that the distance scale used here
is that established by \citet{sarajedini99} based on main sequence 
fitting to open clusters (Section 3). Both of these distance moduli
are consistent with the latest determination based on a combination
of red clump stars, RR Lyraes, the tip of the RGB, and Cepheids
presented by \citet{udalski00}. He finds $(m-M)_0 = 18.75 \pm 0.07$
for the SMC. This agreement underscores the fact that the distance
scale which yields consistent results from these four indicators
is in turn consistent with the open cluster scale of \citet{sarajedini99}.

With the line-of-sight distance to the
clusters and accurate positions in right ascension
and declination \citep{welch91}, we are able to determine the
distribution of these clusters in three dimensional space. 
While the small number of clusters available to us represents a
limitation in studying the overall structure of the SMC, it does allow 
us to examine the principal focus of this work, which is the overall 
depth of the SMC (Section~\ref{depth}). 

\subsection{Correlations Between Observed and Derived Quantities}

As an initial examination of the observational data, Fig. 7 shows
the extinction-corrected
red clump magnitudes for our 12 clusters as a function of their
metal abundances and ages. The former has been calculated using 
$V(RC)$ along with the extinction values inferred from both the BH and
SFD reddening maps. Taking into account the
selection effects and the small number of clusters, Fig. 7 displays
no apparent trends.

We can also examine the behavior of the variation of cluster age and 
metal abundance as a function of position on the sky as shown in Fig. 8. 
Note that Kron 28 has been plotted with the metallicity
inferred by its BH reddening.
There are no definite trends apparent in Fig. 8, and, given the selection
effects and the small number of clusters, it is difficult to 
unequivocally isolate a specific variation; however, the data do hint 
at a possible 
correlation. There is a tendency for the clusters to be younger and
more metal-rich on the eastern side of the SMC (facing the LMC)
as compared with the western side. This phenomenon was also noted
by \citet{gardiner92}; they linked it to the past interactions between
the SMC and LMC. However, \citet{gardiner92} speculate that, instead of 
stimulating star formation in this region,  the interaction 
with the LMC `pulled out' this region
from the SMC exposing its existing stellar populations 
\citep{piatti01}.

Given the cluster distances in Table 3, Fig. 9 illustrates the 
behavior of this distance with position on the sky. Here we have computed
the angular offsets in Right Ascension and Declination from the optical
center of the SMC (Sec. 1) and converted this to a linear distance 
using the distance to each cluster. Once again, the small number of
clusters prevents us from making any definitive
statements. However, it is important to point out that several previous
studies based on the positions of SMC field stars - \citet{hatz89b}
for red clump stars, \citet{mathewson86, mathewson88} for Cepheids,
and \citet{kunkel00} for carbon stars - have concluded that the
northeastern portion of the SMC is closer to us while the southwestern
region is farther away. In contrast, as shown in Fig. 9, the 12 SMC
clusters 
studied herein display little or no evidence of this inclination.
If this result is not simply due to the small number of clusters and 
continues to hold up as more clusters are added to the database, it 
indicates that the SMC field stars do not share in the spatial properties 
of the clusters. The extant kinematical information \citep{dacosta98}
suggests that both of these components possess similar velocity 
characteristics, i.e. no significant rotation. In any case, there appears
to be a paucity of observational data - both for the clusters and field 
stars - to allow an adequate treatment of this apparent discrepancy.

\subsection{The Depth of the Populous Clusters}
\label{depth}

To estimate the line of sight depth of the SMC, we consider the 
standard deviation ($\sigma_{obs}$) of the cluster 
distances using the small-sample statistical formulae of
\citet{keeping62}. 
We note that  $\sigma_{obs}$ is influenced by two factors:
1) The intrinsic depth of the clusters along the line of sight 
($\sigma_{int}$), 
and 2) the fact that each of our cluster distances
is also affected by measurement errors ($\sigma_{err}$). These two factors
combine in quadrature to create the overall
observed standard deviation:
\begin{equation}
\sigma^2_{obs}=\sigma^2_{int}+\sigma^2_{err}.
\end{equation}

\noindent The intrinsic standard deviation is then \(\sqrt{\sigma^2_{obs}
-
  \sigma^2_{err}}\). 

In Table 4, we present the results of this analysis under the four 
different assumptions outlined above. First, if the red clump magnitude 
is sensitive to age and abundance, then, using the curves illustrated in
Fig. 6 and the BH reddenings ($M_V(RC)$ + BH),
we find an intrinsic 1-$\sigma$ depth of $\sigma_{int} = 3.2 \pm 0.7$ kpc.
On the other hand, if we adopt the SFD reddenings
($M_V(RC)$ + SFD), a 1-$\sigma$ depth of $\sigma_{int} = 3.7 \pm 0.8$ kpc 
results. If the red clump magnitude
is insensitive to age and metallicity, then, using the mean 
$M_V(RC)$ of our clusters along with the BH reddenings 
($\langle$$M_V(RC)$$\rangle$ + BH) produces an 
intrinsic SMC depth of $\sigma_{int} = 3.9 \pm 0.8$ kpc. This result 
is modified significantly if we instead utilize the SFD reddenings 
($\langle$$M_V(RC)$$\rangle$ + SFD), in which case 
$\sigma_{int} = 6.0 \pm 1.2$ kpc. 

Therefore, regardless of which assumption we make for the absolute
magnitude
of the red clump, the statistically significant range of distances 
among the populous clusters indicates that the SMC does indeed
exhibit a substantial extent in the LOS direction in agreement with 
several previous investigations. 
Using the language of \citet{gardiner91}, our data indicate
that the SMC has a mean ``2-\(\sigma\) depth'' (i.e. $\pm1\sigma$) of
between \(\sim 6\) kpc and \(\sim 12\) kpc. This is consistent with 
the values quoted by \citet{gardiner91}, which lie anywhere between 
4 and 16 kpc depending on which portion of the SMC one is observing. 

There is one potentially important caveat to keep in mind when
comparing our LOS depth results to those of \citet{hatz89a} and
\citet{gardiner91}. These studies calculated the LOS depth 
by observing several relatively small field regions distributed around
the SMC and using the height of the red clump (corrected for photometric 
errors) as a depth indicator. The overall depth is then the numerical
average of these individual depths. Even though Fig. 9 does not suggest
the
existence of an inclination among the SMC clusters, if there is a
significant 
global inclination to the SMC (field stars and clusters) relative to the
line 
of sight, then our LOS depth as calculated using equation 1 will be 
systematically larger than those quoted by \citet{hatz89a} 
and \citet{gardiner91}.  

To give the reader some `perspective' on the distribution in space
of the SMC clusters, Figs. 10a and b show a three-dimensional plot of the
12 star clusters in the present study under assumptions (1) and (2),
respectively. Figures 9 and 10 suggest that the SMC appears to be a 
triaxial galaxy.
Adopting the Declination, Right Ascension, and LOS depth as the
three axes, we find axial ratios of approximately 1:2:4. 

\section{Conclusions}

In this work, we have compiled published photometry for 12
Small Magellanic Cloud populous clusters that possess helium burning
red clumps. After adopting metallicities on the \citet{zw84} scale
and ages on a scale where Lindsay 1 is 9 Gyr old, we have analyzed
the variation of these properties with position on the sky and
with line-of-sight depth. Based on this analysis, we draw the
following conclusions.

\noindent 1) The observational data indicate that the eastern side
of the SMC (facing the LMC) contains younger and more metal-rich
clusters as compared with the western side. This is not a strong
correlation because our dataset of clusters is necessarily limited,
but it is suggestive and warrants further study.

\noindent 2) Using theoretical models describing the luminosity
of the red clump and its variation with age and abundance zeropointed to
the scale of main sequence fitting distances to Galactic open clusters,
coupled with the Burstein \& Heiles (1982) reddening maps, we find a
mean distance modulus for our 12 SMC clusters of 
$(m-M)_0 = 18.82 \pm 0.05$. If the Schlegel et al. (1998) reddenings
are adopted instead, this is reduced to $(m-M)_0 = 18.71 \pm 0.06$.
These results do not change significantly if the corrections to the
\citet{girardi00} models are applied.

\noindent 3) The intrinsic $\pm$1-$\sigma$ line-of-sight depth of the SMC
populous clusters in our study lies between \(\sim 6\) kpc and
\(\sim 12\) kpc depending on assumptions regarding the variation of
red clump luminosity and the amount of interstellar reddening. 

\noindent 4) Viewing the SMC as a triaxial galaxy with the Declination, 
Right Ascension, and LOS depth as the three axes, we find axial ratios 
of approximately 1:2:4.

Taken together, these conclusions largely agree with those of
previous investigators and serve to underscore the utility of populous
star clusters as probes of the structure of the Small Magellanic Cloud.

\acknowledgements

D.G. acknowledges financial support for this project received 
from CONICYT 
through Fondecyt grants 1000319 and 8000002,
and by the Universidad de Concepci\'on through
research grant No. 99.011.025.

\break

\break

\begin{center}
\Large{Figure Captions}
\end{center}

\figcaption{Mosaic of Digital Sky Survey images obtained from Skyview
  of the twelve clusters and their positions within the SMC.}

\figcaption{Color Magnitude Diagrams of three SMC populous clusters
  with photometry in the Washington Photometric system from the work
  of \citet{piatti01}. A box is drawn around the red
  clump of each cluster.}

\figcaption{NGC 152 CMDs from the four individual CCDs. Notice the
  prominant red clump in planetary camera, which is less obvious in
  the three wide field images. A box is drawn around
  the cluster red clump in each panel.}

\figcaption{Same as Fig. 3 except that the CMDs of NGC 411 are shown.}

\figcaption{Combined Wide Field and Planetary Camera color magnitude 
diagrams of two SMC clusters from observations with the Hubble Space 
Telescope. A box is drawn around the red clump of each cluster.}

\figcaption{The absolute red clump magnitude from the
\citet{girardi00} isochrones zeropointed to the mean 
absolute magnitude of open cluster red clumps from the study of
Sarajedini (1999) at their mean metallicities and ages as shown by
the plotted data point. The models indicated by solid lines are for ages
of (top to bottom)
2.0, 2.5, 5.0, 7.0, 7.9, 10.0, and 12.6 Gyr. The dashed line is for an
age of 1 Gyr.}

\figcaption{The variation of the extinction corrected V magnitude
of the red clump as a function of cluster metallicity (left) and
age (right). The upper panels show the variation when the 
Burstein \& Heiles (1982) reddenings are adopted while the lower
panels how the result of adopting the Schlegel et al. (1998) 
reddenings.}

\figcaption{The variation of cluster age (bottom) and metal abundance
(top) 
as a function of Right Ascension (left) and Declination (right).}

\figcaption{(top) The distance of each cluster in our dataset as a
function
of its Right Ascension (top) and Declination (bottom)
offset from the optical center of the SMC. The left panels show the 
result of adopting the Burstein \& Heiles (1982) reddenings while the
right
panels how the result of the Schlegel et al. (1998) 
reddenings. In the upper panels, east is to the left and in the lower
panels, south is to the left.}

\figcaption{(a) Three dimensional plot of SMC cluster positions assuming
that the red clump magnitude is sensitive to age and abundance along with
the \citet{schelgel98} reddenings. (b) Same as (a) except along a
different line-of-sight.}

\break

\begin{deluxetable}{c c c c}

\tablewidth{3.8in}
\tablecaption{HST Observation Log}
\tablehead{
\colhead{Cluster} & \colhead{Date} &\colhead{Filter} &
\colhead{Exposure Time} \\
\colhead{} & \colhead{(yy/mm/dd)} & \colhead{} & \colhead{(s)}}
\tablecolumns{4}

\startdata
NGC 411 & 94/05/25 & F450W & 400.0 \\
& 94/05/25 & F555W & 200.0 \\
NGC 152 & 94/09/27 & F450W & 300.0 \\
& 94/09/27 & F555W & 160.0 \\
\enddata
\label{table:HSTobs}

\end{deluxetable}

\begin{deluxetable}{l c c c c c c}
\tablewidth{7.5in}
\tablecaption{SMC Cluster Properties}
\tablehead{
\colhead{Cluster} & \colhead{[Fe/H]} &\colhead{Age} &
\colhead{Reference\(^a\)} & \colhead{$V(RC)$} & 
\colhead{$E(B-V)_{BH}^b$} & \colhead{$E(B-V)_{SFD}^b$} \\
\colhead{} & \colhead{} & \colhead{(Gyr)} & \colhead{} & \colhead{} &
\colhead{} & \colhead{}}

\tablecolumns{7}

\startdata
NGC 411 & \(-0.68 \pm 0.07\) & \(1.4 \pm 0.2\) & 1 & \(19.43 \pm 0.05\) &
0.03 & 0.12 \\
NGC 152 & \(-0.94 \pm 0.15\) & \(1.4 \pm 0.2\) & 2 & \(19.57 \pm 0.05\) &
0.03 & 0.05 \\
Lindsay 113 & \(-1.24 \pm 0.11\) & \(5.3 \pm 1.3\) & 3 & \(19.15 \pm
0.02\) & 0.02 & 0.05 \\
Kron 3 & \(-1.16 \pm 0.09\) & \(6.0 \pm 1.3\) & 3 & \(19.45 \pm 0.05\) &
0.02 & 0.03 \\
NGC 339 & \(-1.50 \pm 0.14\) & \(6.3 \pm 1.3\) & 3 & \(19.46 \pm 0.05\) &
0.03 & 0.05 \\
NGC 416 & \(-1.44 \pm 0.12\) & \(6.9 \pm 1.1\) & 3 & \(19.74 \pm 0.05\) &
0.04 & 0.12 \\
NGC 361 & \(-1.45 \pm 0.11\) & \(8.1 \pm 1.2\) & 3 & \(19.53 \pm 0.05\) &
0.05 & 0.13 \\
Lindsay 1 & \(-1.35 \pm 0.08\) & \(9.0 \pm 1.0\) & 3 & \(19.34 \pm 0.02\)
& 0.01 & 0.03 \\
NGC 121 & \(-1.71 \pm 0.10\) & \(11.9 \pm 1.3\) & 3 & \(19.73 \pm 0.05\) &
0.04 & 0.03 \\
Kron 28 & \(-1.20(-1.45) \pm 0.13^c\) & \(2.1 \pm 0.5\) & 4 & \(19.33 \pm
0.05\) & 0.06 & 0.16 \\
Lindsay 38 & \(-1.65 \pm 0.12\) & \(6.0 \pm 0.5\) & 4 & \(19.56 \pm 0.05\)
& 0.02 & 0.02 \\
Kron 44 & \(-1.10 \pm 0.11\) & \(3.1 \pm 0.8\) & 4 & \(19.52 \pm 0.05\) &
0.02 & 0.05 \\

\enddata

\tablenotetext{a}{(1) \citet{alves99}; (2) This paper; (3)
 \citet{mighell98}; (4) \citet{piatti01}}
\tablenotetext{b}{An error of 10\% has been assumed in the reddenings.}
\tablenotetext{c}{As determined by Piatti et al. (2000) adopting a 
reddening from Burstein \& Heiles (1982). The value in parenthesis is the 
result of adopting the Schlegel et al. (1998) reddening.}
\label{table:agemetal}
\end{deluxetable}

\begin{deluxetable}{l c c c}
\tablewidth{5.in}
\tablecaption{SMC Cluster Red Clump Distances}
\tablehead{
\colhead{Cluster} & \colhead{\(M_V(RC)\)} & \colhead{Distance(BH)$^a$} & 
\colhead{Distance(SFD)$^b$}\\
\colhead{} & \colhead{} & \colhead{(kpc)} & \colhead{(kpc)}}
\tablecolumns{4}

\startdata
NGC 411 & \(0.63 \pm 0.16\) & \(55.1 \pm 4.3\) & \(48.5 \pm 3.8\)\\
        & \(0.56 \pm 0.04\) & \(56.9 \pm 1.6\) & \(50.1 \pm 1.7\)\\
        &                   &                  &                 \\
NGC 152 & \(0.49 \pm 0.19\) & \(62.7 \pm 5.7\) & \(61.0 \pm 5.5\)\\
        & \(0.56 \pm 0.04\) & \(60.7 \pm 1.9\) & \(59.0 \pm 1.8\)\\
        &                   &                  &                 \\
Lindsay 113 & \(0.57 \pm 0.07\) & \(50.5 \pm 1.7\) & \(48.4 \pm 1.7\)\\ 
            & \(0.56 \pm 0.04\) & \(50.8 \pm 1.1\) & \(48.6 \pm 1.1\)\\
        &                   &                  &                 \\
Kron 3 & \(0.62 \pm 0.05\) & \(56.7 \pm 1.9\) & \(55.9 \pm 1.8\)\\
       & \(0.56 \pm 0.04\) & \(58.3 \pm 1.7\) & \(57.5 \pm 1.7\)\\
        &                   &                  &                 \\
NGC 339 & \(0.57 \pm 0.06\) & \(57.5 \pm 2.1\) & \(55.8 \pm 2.0\)\\
        & \(0.56 \pm 0.04\) & \(57.7 \pm 1.7\) & \(56.1 \pm 1.7\)\\
        &                   &                  &                 \\
NGC 416 & \(0.60 \pm 0.05\) & \(63.6 \pm 2.1\) & \(56.7 \pm 2.1\)\\
        & \(0.56 \pm 0.04\) & \(64.7 \pm 1.9\) & \(57.8 \pm 2.0\)\\
        &                   &                  &                 \\
NGC 361 & \(0.64 \pm 0.04\) & \(55.8 \pm 1.7\) & \(49.8 \pm 1.7\)\\
        & \(0.56 \pm 0.04\) & \(57.9 \pm 1.7\) & \(51.7 \pm 1.8\)\\
        &                   &                  &                 \\
Lindsay 1 & \(0.68 \pm 0.03\) & \(53.2 \pm 0.9\) & \(51.7 \pm 0.9\)\\
          & \(0.56 \pm 0.04\) & \(56.2 \pm 1.1\) & \(54.6 \pm 1.2\)\\
        &                   &                  &                 \\
NGC 121 & \(0.73 \pm 0.04\) & \(59.6 \pm 1.8\) & \(60.5 \pm 1.8\)\\
        & \(0.56 \pm 0.04\) & \(64.4 \pm 1.8\) & \(65.4 \pm 1.9\)\\
        &                   &                  &                 \\
Kron 28 & \(0.27 \pm 0.12\) & \(59.5 \pm 3.6\) & \(54.8 \pm 3.3\)\\
        & ($0.14\pm0.11$)$^c$ &                &                 \\
        & \(0.56 \pm 0.04\) & \(52.1 \pm 1.8\) & \(45.2 \pm 1.7\)\\
        &                   &                  &                 \\
Lindsay 38 & \(0.51 \pm 0.03\) & \(62.7 \pm 1.9\) & \(62.7 \pm 1.9\)\\
           & \(0.56 \pm 0.04\) & \(61.3 \pm 1.9\) & \(61.3 \pm 1.8\)\\
        &                   &                  &                 \\
Kron 44 & \(0.49 \pm 0.08\) & \(62.2 \pm 2.7\) & \(59.6 \pm 2.6\)\\
        & \(0.56 \pm 0.04\) & \(60.2 \pm 1.8\) & \(57.7 \pm 1.8\)\\
\enddata

\tablenotetext{a}{Calculated adopting the red clump absolute magnitudes
in Column (2) and the
Burstein \& Heiles (1982) reddenings listed in column 5 of Table 2.}
\tablenotetext{b}{Calculated adopting the red clump absolute magnitudes
in Column 2 and the
Schlegel et al. (1998) reddenings listed in column 6 of Table 2.}
\tablenotetext{c}{As determined from Fig. 6 adopting the metallicity from
Piatti et al. (2000) derived assuming a 
reddening from Burstein \& Heiles (1982). The value in parenthesis is the 
result of adopting the metallicity yielded by the Schlegel et al. (1998)
reddening
(see Table 2).}
\label{table:mags} 
\end{deluxetable}

\begin{deluxetable}{l c c c}
\tablewidth{4.in}
\tablecaption{SMC Depth Results}
\tablehead{
\colhead{Assumption} & \colhead{$\sigma$$_{obs}$} & 
\colhead{$\sigma$$_{err}$} & \colhead{$\sigma$$_{int}$}\\
\colhead{} & \colhead{kpc} & \colhead{(kpc)} & \colhead{(kpc)}}
\tablecolumns{4}

\startdata
1) $M_V(RC)$ + BH & $4.1 \pm 0.8$ & $2.5 \pm 0.4$ & 
$3.2 \pm 1.0$ \\
2) $M_V(RC)$ + SFD & $4.4 \pm 0.9$ & $2.4 \pm 0.4$ & 
$3.7 \pm 1.1$ \\
3) $\langle$$M_V(RC)$$\rangle$ + BH & $4.3 \pm 0.9$ & $1.7 \pm 0.1$ & 
$3.9 \pm 1.0$ \\
4) $\langle$$M_V(RC)$$\rangle$ + SFD & $6.3 \pm 1.3$ & $1.7 \pm 0.1$ & 
$6.0 \pm 1.7$ \\
\enddata
\end{deluxetable}


\begin{thebibliography}

\bibitem[Alves \& Sarajedini(1999)]{alves99} Alves, D. \& Sarajedini, A.
1999, \apj, 511, 225

\bibitem[Bertelli et al.(1994)]{bertelli94} Bertelli, G., Bressan, A., 
  Chiosi, C., Fagotto, F., \& Nasi, E., 1994, A\&AS, 106, 275

\bibitem[Bica \& Schmitt(1995)]{bica95}
Bica, E., \& Schmitt, H. R. 1995, \apjs, 101, 41

\bibitem[Bica et al.(1999)]{bica99}
Bica, E., Schmitt, H. R., Dutra, C. M., \& Oliveira, H. L. 1999, \aj, 117,
238

\bibitem[Burstein \& Heiles(1982)]{bh}
Burstein, D., \& Heiles, C. 1982, \aj, 87, 1165

\bibitem[Canterna(1976)]{canterna76} Canterna, R. 1976, \aj, 81, 228

\bibitem[Cardelli et al.(1989)]{cardelli89}
Cardelli, J. A., Clayton, G. C., \& Mathis, J. S. 1989, \apj, 345, 245

\bibitem[Carlson et al.(1998)]{carlson98}
Carlson, M. N., Holtzman, J. A., Watson, A. M., Grillmair, C. J.,
Mould, J. R., Ballester, G. E., Burrows, C. J., Clarke, J. T., 
Crisp, D., Evans, R. W., Gallagher, J. S., Griffiths, R. E.,
Hester, J. J., Hoessel, J. G., Scowen, P. A., Stapelfeldt, K. R.,
Trauger, J. T., \& Westphal, J. A. 1998, \aj, 115, 1778

\bibitem[Castellani et al.(2000)]{castellani00} Castellani, V.,
  Degl'Innocenti, S., Girardi, L., Marconi, M., Prada Moroni, P. G.,
  Weiss, A., A\&A, 354, 150

\bibitem[Cole(1998)]{cole98}
Cole, A. 1998, ApJ, 500, L137

\bibitem[C\^{o}t\'{e} et al.(2000)]{cote00} C\^{o}t\'{e}, P., Marzke,
  R.O., West, M.J., \& Minniti, D. 2000, \apj, 533, 869 


\bibitem[Da Costa \& Hatzidimitriou(1998)]{dacosta98}
Da Costa, G. S., \& Hatzidimitriou, D. 1998, \aj, 115, 1934
  
  \bibitem[Da Costa et al.(1987)]{dkm} Da Costa, G. S., King, C. R.,
\& Mould, J. R. 1987, \apj, 321, 735

\bibitem[Elmegreen et al.(2000)]{elmegreen00}
Elmegreen, B. G., Kaufman, M., Struck, C., Elmegreen, D. M., Brinks, E., 
Thomasson, M., Klaric, M., Levay, Z., English, J., Frattare, L. M.,
Bond, H. E., Christian, C. A., Hamilton, F., \& Noll, K. 2000, \aj, 
120, 630

\bibitem[Gardiner(1999)]{gardiner99}
Gardiner, L. T. 1999, in New Views of the Magellanic Clouds, IAU
Symp. No. 190, (ASP: San Francisco) p. 480

\bibitem[Gardiner \& Hawkins(1991)]{gardiner91} Gardiner, L.T. \&
  Hawkins, M.R.S. 1991, \mnras, 251, 174

\bibitem[Gardiner \& Hatzidimitriou(1992)]{gardiner92} Gardiner, L.T. \&
  Hatzidimitriou, D. 1992, \mnras, 257, 195

\bibitem[Geisler(1996)]{geisler96} Geisler, D. 1996, \aj, 111, 480

\bibitem[Geisler \& Sarajedini(1999)]{geisler99}
Geisler, D., \& Sarajedini, A. 1999, \aj, 117, 308

\bibitem[Geisler et al.(1997)]{geisler97} Geisler, D., Bica, E., 
Dottori, H., Clari\'{a}, J. J., Piatti, A. E., \& Santos, J. F. C., Jr.
1997, AJ, 114, 1920

\bibitem[Girardi(1999)]{girardi99} Girardi, L. 1999, MNRAS, 308, 818

\bibitem[Girardi et al.(2000)]{gmc00} Girardi, L, Mermilliod,
  J.-C., Carraro, G. 2000, A\&A, 354, 892

\bibitem[Girardi \& Salaris(2000)]{salaris00}
Girardi, L., \& Salaris, M. 2000, MNRAS, in press (astro-ph/0007343)

\bibitem[Girardi et al.(2000)]{girardi00} Girardi, L., Bressan, A.,
  Bertelli, G., \& Chiosi, C. 2000, A\&AS, 141, 371 

\bibitem[Hatzidimitriou et al.(1989a)]{hatz89a} 
Hatzidimitriou, D., Hawkins, M.R.S., Gyldenkerne, K. 1989a, \mnras, 241,
645

\bibitem[Hatzidimitriou et al.(1989b)]{hatz89b} Hatzidimitriou, D.,
  Hawkins, M.R.S. 1989b, \mnras, 241, 667

\bibitem[Holtzman et al.(1995)]{holtzman95} Holtzman, J.A., Burrows, C.J.,
 Casertano, S., Hester, J.J., Trauger, J.T., Watson, A.M., \& Worthey,
 G. 1995, \pasp, 107, 156

\bibitem[Keeping(1962)]{keeping62} Keeping, E.S. 1962,
\underline{Introduction
  to Statistical Inference}, (Princeton: Van Nostrand)

\bibitem[Kunkel et al.(2000)]{kunkel00} Kunkel, W.E., Demers, S., \&
  Irwin, M.J. 2000, \aj, 119, 2789

\bibitem[Mathewson(1984)]{mathewson84} Mathewson, D.S. 1984,
  \emph{Mercury}, March-April, 57

\bibitem[Mathewson et al.(1986)]{mathewson86} Mathewson, D.S., Ford,
  V.L., Visvanathan, N. 1986, \apj, 301, 664

\bibitem[Mathewson et al.(1988)]{mathewson88} Mathewson, D.S., Ford,
  V.L., Visvanathan, N. 1988, \apj, 333, 617

\bibitem[Mighell et al.(1998)]{mighell98} Mighell, K.J., Sarajedini,
  A., \& French, R.S. 1998, \aj, 116, 2395 

\bibitem[Murai \& Fujimoto(1980)]{murai80} Murai, T., \& Fujimoto, M. 
  1980, \pasj, 32, 581

\bibitem[Noriega-Mendoza \& Ruelas-Mayorga(1997)]{noriega97}
Noriega-Mendoza, H., \& Ruelas-Mayorga, A. 1997, \aj, 113, 722

\bibitem[Olszewski et al.(1996)]{osm96}
Olszewski, E. W., Suntzeff, N., \& Mateo, M. 1996, ARA\&A, 34, 511

\bibitem[Paczy\'{n}ski \& Stanek(1998)]{paczynski98}
Paczy\'{n}ski, B., \& Stanek, K. Z. 1998, \apj, 494, L219

H.

\bibitem[Piatti et al.(2001)]{piatti01} Piatti, A.E., Santos Jr.,
  J.F.C., Clari\'{a}, J.J., Bica, E., Sarajedini, A., Geisler,
  D. 2000, \aj, submitted 

\bibitem[Rich et al.(2000)]{rich00} Rich, R.M., Shara, M., Fall, S.M.,
\& Zurek, D. 2000, \aj, 119, 197

\bibitem[Sarajedini(1994)]{sarajedini94} Sarajedini, A. 1994, \aj,
  107, 618

\bibitem[Sarajedini(1998)]{sarajedini98} Sarajedini, A. 1998, \aj,
  116, 738

\bibitem[Sarajedini(1999)]{sarajedini99} Sarajedini, A. 1999, \aj,
118, 2321

\bibitem[Sarajedini, Lee, \& Lee(1995)]{sll95}Sarajedini, A., Lee, Y.-W.,
\& 
  Lee, D.-H. 1995, \apj, 450, 712

\bibitem[Searle \& Zinn(1978)]{searle78}Searle, L. \& Zinn, R. 1978,
  \apj, 225, 357

\bibitem[Schlegel et al.(1998)]{schlegel98} Schlegel, D.J.,
  Finkbeiner, D.P., \& Davis, M. 1998, \apj, 500, 525 

\bibitem[Stetson(1994)]{stetson94} Stetson, P.B. 1994, \pasp, 106, 250

\bibitem[Udalski(1998)]{udalski98}
Udalski, A. 1998, AcA, 48, 383

\bibitem[Udalski(2000)]{udalski00}
Udalski, A. 2000, AcA, 50, 279

\bibitem[Wannier \& Wrixon(1972)]{wannier72} Wannier, P. \& Wrixon,
  G.T. 1972, \apj, 119, L119

\bibitem[Welch et al.(1987)]{welch87} Welch, D.L., McLaren, R.A.,
  Madore, B.F., \& McAlary, C.W. 1987 \apj, 321, 162

\bibitem[Welch(1991)]{welch91} Welch, D.L. 1991, \aj, 101, 538

\bibitem[Westerlund(1990)]{westerlund90}
Westerlund, B. E. 1990, A\&ApR, 2, 29

\bibitem[Whitmore et al.(1999)]{whitmore99}
Whitmore, B. C., Zhang, Q., Leitherer, C., Fall, S. M., Schweizer, F.,
\& Miller, B. W. 1999, \aj, 118, 1551

\bibitem[Zinn \& West(1984)]{zw84} Zinn, R. J., \& West, M. J. 1984,
\apjs, 55, 45

\end{thebibliography}
\end{document}